# UNRAVELLING DNS PERFORMANCE: A HISTORICAL EXAMINATION OF F-ROOT IN SOUTHEAST ASIA

Jiajia Zhu and Chao Qi

China Academy of Information and Communications Technology, Beijing, China

## ABSTRACT

*The DNS root server system uses Anycast technology to provide resolution through widely distributed root nodes. In recent years, the F-root node has seen astonishing growth and now boasts the largest number of nodes among the 13 root servers. Based on Ripe Atlas measurement data, we examined the availability and query latency of the F-root within the Southeast Asian region historically. The collected data illustrates how latency varies with changes in the number of root nodes, how the geographic distribution of responding root nodes changes in different periods, and examines the most recent differences between countries in terms of latency distribution. This study sheds light on the evolving landscape of DNS infrastructure in Southeast Asia.*

## KEYWORDS

*DNS, Southeast Asia, Ripe Atlas, Query Latency*

## 1. INTRODUCTION

Root server is the foundational component of Domain Name System (DNS) and a critical element of the internet infrastructure [1]. Presently, there are 12 Internet Corporation for Assigned Names and Numbers (ICANN) authorized root operating organizations responsible for managing a total of 13 root name servers, denoted from A-root to M-root. To safeguard against Distributed Denial of Service (DDoS) attacks [2] on the DNS root and to minimize root query latency, the Root Server System (RSS) deploys numerous mirrored nodes (replicating instances of root service) at different sites worldwide, employing Anycast technology [3,4]. Notably, the number of F-root nodes has surged from 58 in early 2016 to 505 as of August 2023, marking a nearly tenfold increase. This deployment makes it the fastest-growing and the largest in terms of the root nodes number among A-M roots. Among all continents, Asia leads the way, with the number of F-root nodes growing from 15 in early 2016 to 191 as of August 2023 [5].

Extensive research has been conducted around DNS in the past, including topics around root server deployment [6,7], root traffic [8], and root manipulation [9]. As the deployment process of root nodes accelerates, RSS has undergone great changes in recent years, so it is necessary to conduct a fresh round of examination of the current RSS. Among the research after 2020, some focus on the intersection of DNS and new technologies such as blockchain [10,11], some delve into DNS security and privacy [12,13,14], and some address the centralization challenges [15] within DNS ecosystem. However, there are few studies on root service performance, and even fewer studies on DNS root performance of a certain region from a historical perspective.

This article centres its focus on the F-root and selects Southeast Asia ("SEA" for short) as the primary research area. Leveraging data from Ripe Atlas [16,17] and the root-servers.org platform, alongside considerations of the number of Ripe probes and the availability of historical data, six countries in SEA have been chosen as the subjects of this study. These countries include Indonesia, Malaysia, the Philippines, Singapore, Thailand, and Vietnam. The subsequent structure of this paper is as follows:

Section 2 outlines the research methodology. Section 3 provides an in-depth analysis of the research findings, encompassing a historical examination of the deployment and query latency of the F-root within the SEA region and individual countries. Additionally, it delves into the changing geographical distribution of destination nodes and presents the most recent latency Cumulative Distribution Function (CDF) curves for various countries when accessing the F-root. Section 4 summarizes the research and offers corresponding recommendations.

## 2. METHODOLOGY

### 2.1. Research Scope

Building on the findings of a report from LACNIC [18], this study represents the first long-term analysis focusing on the operational status of DNS root in the Southeast Asian region. Data regarding the query performance of the F-root in this study is sourced from the Ripe Atlas platform. Ripe Atlas is a global probe network designed for testing network connectivity and reachability, providing real-time insights into network conditions. Currently, Ripe Atlas consists of more than 12000 of active probes distributed worldwide, with the number continuing to grow.
Among all the SEA countries, we selected those that have had active probes since early 2016 and, as of August 2023, have at least 10 active probes. Meeting these criteria are only the aforementioned six countries. While they do not represent the entirety of the SEA region, these six countries collectively account for 88% of the region's population and 75% of its territorial area. Several other SEA countries were unable to be included due to no probes or an insufficient probe count. Thus, this study approximates the overall situation in the SEA region by focusing on these six countries.

### 2.2. Data Collection

DNS measurements are built-in measurements of the Ripe platform and all data are publicly accessible [19]. In this study, we collected DNS measurement records for the F-root from all probes in the six countries mentioned, sampled at regular time intervals from Jan 2016 to Aug 2023. Specifically, the measurements are IPv4 DNS queries for QNAME "hostname.bind" over UDP, using CHAOS class and TXT type [20]. This allowed us to gather information about the response time and the hostnames of the responding root node from each record.

In addition to RIPE's data, this study also utilized historical archive data from the root-servers.org platform regarding the deployment of root servers, such as information about root node types and geographical locations.

### 2.3. Root Node's Identification and Location

Several articles have previously discussed methods for identifying and locating root nodes [21,22,23]. In this paper, we employ a relatively simple and straightforward method. The process can be summarized as follows:

—We obtain the hostname of the responding root node from RIPE's DNS measurement records, found in the "hostname.bind" field.

—We use this hostname to match it with the "identifier" field in archive yaml/json data from root-servers.org website. Root-servers.org data provides identifiers for the F-root, which makes this matching process possible.

—If the match is successful, it automatically identifies the specific root node that responded in each measurement record and provides its location information.

It's important to note that while this method is effective for the F-root, it may not be applicable to all other 12 root servers because some root servers' identifiers are missing in the data provided by root-servers.org.

## 3. RESULTS

### 3.1. Use of F-Root in Region of Southeast Asia

In this section, we analyze the utilization of F-root nodes in the SEA region, focusing on the number of nodes and query latency. To illustrate this, we utilize the combination chart presented in Fig. 1, which provides insights into query latency to the F-root by various probes at different historical periods. In the chart, the horizontal axis represents time, the main vertical axis represents round-trip time (RTT), and the secondary vertical axis represents the number of F-root nodes. The chart displays two lines: the black line signifies the number of F-root nodes, while the red line represents the average RTT of all probes in the region when querying the F-root at that time. Additionally, scattered data points are visible in blue, where each point signifies the average RTT of an individual probe at a given time, with the shades of colour indicating density. Darker colours indicate a higher concentration of similar RTT results, while lighter colours represent the opposite. Generally, a denser distribution along the x-axis signifies better performance.

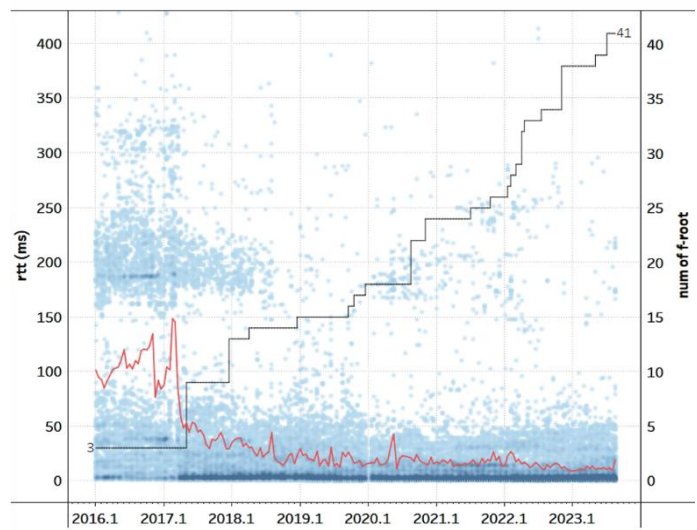

Figure 1.  Historical query latency and number of F-root in SEA region

Our observations reveal that prior to 2016, there were several F-root nodes deployed in the SEA region, although their numbers were relatively low. Before May 2017, some measurement records indicated lower RTT values, situated near the x-axis on the chart. However, a cluster of data points near the upper part of the y-axis was also evident, with a significant concentration around 200ms and a few around 300ms. In May 2017, the region added six new F-root nodes, leading to noticeable changes in the distribution of data points. The number of points at the upper position decreased, and the distribution narrowed. Near the x-axis, the colour became darker, signifying a significant decrease in RTT. By October 2018, the region had 13 root nodes, and scatter points at the upper "layer" became less pronounced and clustered, with only sparse points above. Over time, beyond 2020, the width of the distribution at the lower position gradually narrowed compared to previous periods.

Simultaneously, we observed that while the number of F-root nodes continued to increase after 2020, there were no apparent changes in the distribution of scatter points in Fig. 1. Additionally, the decline in the red curve became increasingly gradual, suggesting that the introduction of a large number of root nodes in the later period did not yield the same degree of improvement as earlier introductions.

## 3.2. Use of F-Root in Each Country of Southeast Asia

In this section, we employ the same research methodology as in the previous section to individually analyze the historical changes in query latency and the number of F-root nodes in six different selected countries within Southeast Asia. The subplots in Fig. 2 present these analyses, with each subplot dedicated to one of the six countries. The country's ISO two-letter code is displayed in the top centre of each subplot.

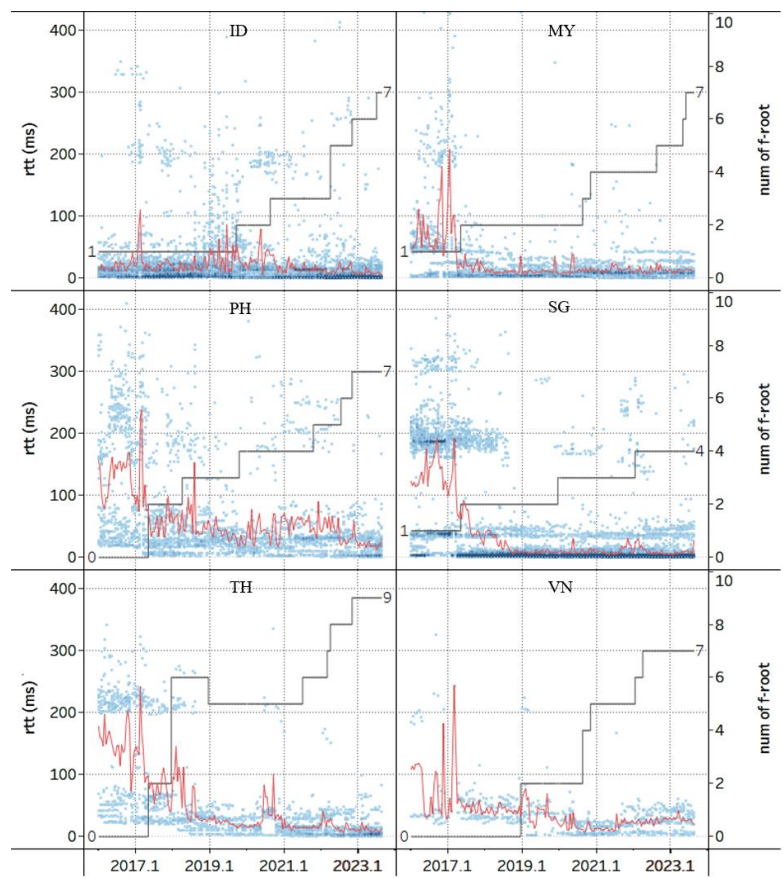

Figure 2.  Historical query latency and number of F-root in each country of SEA

The variations in the number of data points within the subplots of Fig. 2 are attributed to the differing numbers of active probes in each country. As of August 2023, the number of active probes for each country is as follows: Indonesia (92 probes), Malaysia (27 probes), the Philippines (54 probes), Singapore (129 probes), Thailand (34 probes), and Vietnam (10 probes). Consequently, subplots for Indonesia and Singapore exhibit a higher data density, while the subplot for Vietnam features relatively fewer data points.

Observing each country individually, we can find:

**Indonesia (ID)**: Despite an increase in the number of F-root nodes from 1 to 7, query latency for Indonesian probes did not show a clear improvement over time. In the first half of 2016, most data points were already concentrated near the x-axis, with few data points higher up, indicating good latency performance at that time. However, as the number of F-root nodes increased, scatter points at higher position not only did not decrease but, in some instances, showed an increase and even a tendency to cluster in a small range.

**Malaysia (MY)**: Although Malaysia initially had one root node, scatter points were still distributed at various heights along the y-axis until early 2017, indicating a significant latency range. By April 2017, Malaysia introduced a second F-root node, and it was then clearly observed that the scatter points began to cluster downwards. After 2018, most scatter plots clustered at positions below 50ms. While, further node additions did not yield significant enhancements as before.

**The Philippines (PH)**: Before 2017, the Philippines had no domestic F-root, resulting in a broad distribution of query latencies. In its subplot, we can see many scatter points distributed at different heights. With the introduction of its first F-node in April 2017, latency significantly improved, manifested as a gradual decrease in points at higher positions and an increase in points closer to the x-axis. By 2020, its F-root had increased to 4, and by this point, most of the scatter points were located below 50ms. While, like MY, the subsequent increase in the number of root nodes did not bring continued performance improvements, instead we observed a slight deterioration between 2021 and 2022.

**Singapore (SG)**: Although Singapore had already deployed an F-root node before 2016, until early 2017, along the y-axis direction, scatter points could still be clearly observed to form three layers. Among them, the middle layer had the most significant latency span, with a higher density at around 185ms (indicated by a very dark short line). In May 2017, Singapore introduced a new F-root node, and since then, scatter points in the top and middle layers started to descend. Meanwhile, points in the bottom layer began to cluster more towards the very bottom (closer to the x-axis). By August 2018, the upper and middle layers had almost disappeared, and most of the points were descended in the bottom layer.

**Thailand (TH)**: Basically, trends in Thailand paralleled those in Singapore. Initially, the absence of F-root nodes resulted in broad latency distribution. In May 2017, Thailand introduced 2 F-root nodes, which improved the situation. Points in the upper layer started to decrease, by December 2017, the number of F-root nodes increased to 6, and we observed a continued decrease in points in the upper layer, meanwhile, points in the lower layer started to settle even lower. During the first quarter of 2018, points in the upper layer gradually decreased to near "disappearance", and points in the lower layer sank to the very bottom. Afterward, the F-root kept increasing, but latency did not show significant improvement.

**Vietnam (VN)**: After introducing its first F-root node in December 2018, Viet Nam's latency improved considerably. From its subplot, we can see almost all the scatter points at the higher position disappeared, and points below started to cluster further downward. Its best performance was observed between May 2020 and February 2022, manifested as almost all the points were distributed in a narrow strip at the bottom. However, after February 2022, despite an increase in nodes, there was no continued improvement observed as we expected.

In summary, historical trends in query latency for F-root nodes exhibited variations among these countries. However, a common pattern emerged among all five countries, except for Indonesia. Initially, querying F-root nodes resulted in relatively poor performance, characterized by numerous data points clustering at higher positions in the subplots, and the red line indicating high average RTT at the same time. The introduction of additional root nodes led to a significant reduction or near disappearance of data points at higher positions, and the red line also swiftly declined. Furthermore, this analysis reaffirms the conclusion from the previous section: early introductions of root nodes consistently produced substantial improvements in latency, while late introductions had a more limited impact.

### 3.3. Geographical Distribution of Responding Root

In this section, we examine the historical trends in the distribution of destination countries or regions to which responding F-root nodes belong, based on the location of root nodes. Fig. 3 presents the findings, with each row representing an observed country, numerical labels at the top denoting the years, and the bottom numbers (1-4) indicating the quarters of a year. Different colours represent the countries or regions where responding root nodes are situated.

To enhance clarity, we merged destination countries or regions with proportions below 1% into a single category labelled as 'others.' Additionally, some responding root nodes could not be matched with the F-root's Identifier from root-servers.org and were collectively categorized as 'unknown' in this study.

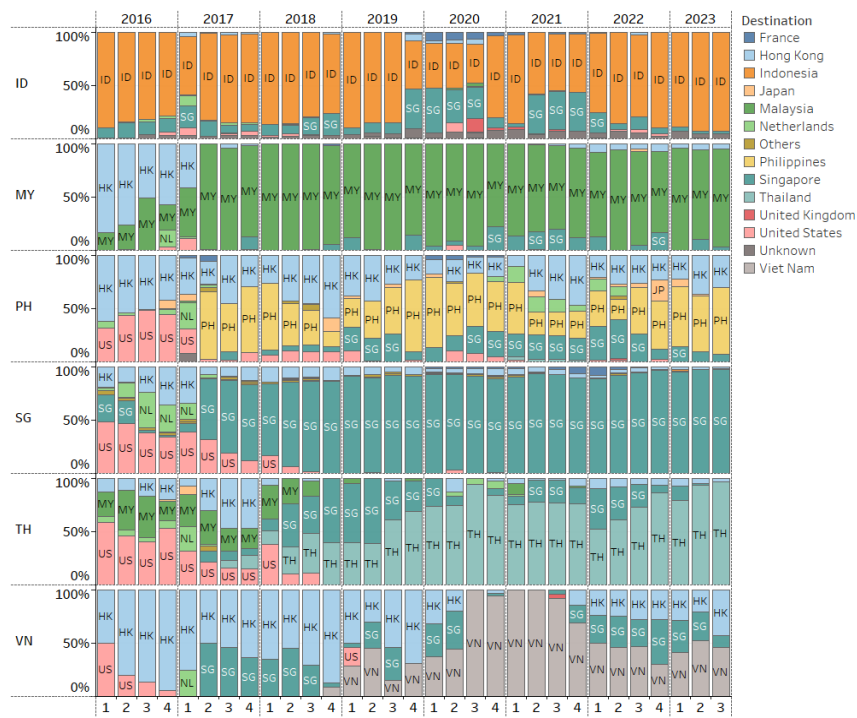

Figure 3.  Geo-distribution of responding F-root servers by country Findings are:

**Regional F-root access**. In Fig. 3, we observe that Singapore's F-root is widely accessed by the entire SEA region. Apart from Singapore, queries from the other five countries also access Singapore's F-root during different periods, some with relatively high proportions. Malaysia's F-root has been accessed by other countries, particularly from 2016 to the third quarter of 2018 when Malaysia was among Thailand's top three destination countries for F-root. On the other hand, Indonesia's and Thailand's F-root nodes were briefly accessed by probes from outside their respective countries, but the proportions were so low that they appear indistinguishable in the graph. Philippines and Vietnam only have 'Local' (with service limited to specific Autonomous Systems) type of F-root, and data shows they are accessible only to probes within their own countries throughout the entire timespan.

**Change in routing destination**. In the early stages, notably in 2016 and 2017, except for Indonesia and Malaysia, the other four countries had a significant portion of their F-root queries routing to the United States or the Netherlands. However, after 2023, none of the six countries has its F-root queries routed to root nodes located outside of Asia any more. The primary destination for F-root in each country is now their own country, with domestic resolution rates exceeding 90% for Indonesia, Malaysia, Singapore, and Thailand. For the Philippines and Viet Nam, these rates was lower at 63% and 46%, respectively, which we speculate that this may be related to the adjustment of their tele-operators' routing policy. Besides their own country, these two countries primarily accessed F-root in Hong Kong and Singapore.

**Impact of introducing more root nodes**. It may be intuitive to think that introducing more F-root nodes would increase the proportion of responses from domestic root nodes. However, our findings reveal that as countries introduced more F-root nodes, the proportion of responses from domestic roots did not consistently increase; instead, it fluctuated. This suggests that countries may not necessarily need so many F-root nodes.

## 3.4. The Latest Latency CDF of Querying F-Root

In this final part of our analysis, we focused on data extracted from July to August of 2023 and plotted cumulative distribution function (CDF) curves of RTT for querying the F-root by each country. In Fig. 4, each line with a distinct shape represents a different country, with ISO two-letter codes providing clarity.

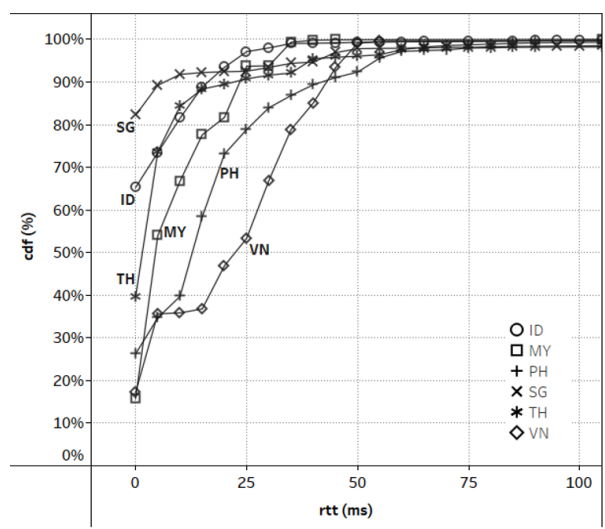

Figure 4. Each country's CDF of RTT

Fig. 4 illustrates that all countries have a proportion of RTT under 50ms exceeding 90%. However, a closer examination of the proportion of RTT under 25ms reveals gaps among countries. Indonesia, Malaysia, Singapore, and Thailand consistently maintain proportions above 90%. In contrast, the Philippines and Vietnam, while achieving good average RTTs (20ms and 24ms, respectively, as seen in Fig. 2), exhibit room for improvement in reducing RTT variances, with proportions under 25ms at 79% and 53%, respectively.

## 4. CONCLUSIONS

In this study, we conducted a historical analysis of use of F-root in the SEA region, focusing on six selected countries and relying on data from the Ripe Atlas and root-servers.org platforms. The results show a clear and continuous evolution towards improved root service performance since 2016.

**1)       Regional improvement**. The introduction of F-root nodes has positively impacted the region, with the most significant improvements occurring in the early period (2017-2018). Indonesia, unique in its good initial latency, didn't experience significant performance enhancements with additional F-root nodes. In contrast, the other five countries followed the regional trend, showing substantial latency improvements with the introduction of root nodes in the early phase. Later additions of root nodes yielded diminishing returns.

**2)       Geographical distribution**. Initially, countries heavily relied on F-root nodes outside its own region, including those in Hong Kong, the United States, and Europe. As F-root nodes increased, each country ended up with querying more of their own root. Presently, Indonesia, Malaysia, Singapore, and Thailand have domestic resolution rates exceeding 90%, while Vietnam and the Philippines lagged with domestic resolution rates of 63% and 46%, respectively.

**3)       RTT CDF characteristics**. Analysis of the latest data from the third quarter of 2023 revealed that over 90% of queries had latency values below 50ms for all countries. However, the Philippines and Vietnam exhibited significant lags when it came to smaller latency values (like under 25ms). Although the current average latency for all countries is favourable, the Philippines and Vietnam can still work to reduce their latency variances.

Given all the findings, it is our view that adding more F-root to these countries appears unnecessary and we advocate for exploring alternative strategies such as implementing a local copy of the root zone file (RFC 8806) [24] or optimizing the routing strategy are worth considering to further optimize the DNS infrastructure in the SEA region.

**Limitations and future work.** This study sheds light on the evolving landscape of DNS infrastructure in SEA region. However, it's crucial to acknowledge the limitations inherent in our study, primarily stemming from the uneven distribution of Ripe probes globally. The limited presence of probes in certain countries, notably in countries of SEA, left us no choice but to focus on only six countries, potentially impacting the generalizability of findings across the entire region. Future research directions should expand beyond the scope of this study. A broader investigation encompassing other root letters beyond F-root would provide a comprehensive understanding of this region. And each specific root node's utilization and the discovered unauthorized root servers highlight an intriguing avenue for further exploration.

**ACKNOWLEDGEMENT**

This work was supported by the Industrial Internet Innovation and Development Project ——
Industrial Internet Identification Resolution System: National Top-Level Node Construction
Project.

**AUTHORS**

**Jiajia Zhu**, a data analyst in CAICT, skilled in data analysis in ICT related filed.

**Chao Qi**, an engineer of R&D and Maintenance in CAICT, skilled in DNS, industrial IOT and Blockchain.